\documentclass[5pt]{article}
\usepackage{amsmath}
\usepackage[latin1]{inputenc}
\usepackage{amsmath,amsthm,amssymb}
\usepackage{graphicx}
\usepackage{color}
\usepackage{multirow}
\usepackage{rotating}

\renewcommand{\theequation}{\thesection.\arabic{equation}}

\title{ Optimal Equi-difference Conflict-avoiding Codes}
\author{{Derong Xie  \qquad\quad    Jinquan Luo}\footnote{ The authors are with school of mathematics and
statistics \& Hubei Key Laboratory of Mathematical Sciences,
Central China Normal University
Wuhan 430079, China.\newline
E-mail: derongxie@yahoo.com(D.Xie), luojinquan@mail.ccnu.edu.cn(J.Luo)
}}
\date{}
\begin{document}
\baselineskip15pt \maketitle
\renewcommand{\theequation}{\arabic{section}.\arabic{equation}}
\catcode`@=11 \@addtoreset{equation}{section} \catcode`@=12

$\mathbf{Abstract-}$An equi-differece conflict-avoiding code $(CAC^{e})\ \mathcal{C}$ of length $n$ and weight $\omega$ is a collection of $\omega$-subsets (called codewords) which has the form $\{0,i,2i,\cdots,(\omega-1)i\}$ of $\mathbb{Z}_{n}$  such that $\Delta(c_{1})\cap\Delta(c_{2})=\emptyset$ holds for any $c_{1},\ c_{2}\in\mathcal{C}$, $c_{1}\neq c_{2}$ where $\Delta(c)=\{j-i \ (\mbox{mod}\ n) \; | \; i,j\in c,i\neq j\}.$ A code $\mathcal{C}\in CAC^{e}s$ with maximum code size for given $n$ and $\omega$ is called optimal and is said to be perfect if $\cup_{c\in \mathcal{C}}\Delta(c)=\mathbb{Z}_{n}\backslash \{0\}.$ In this paper, we show how to combine a $\mathcal{C}_{1}\in CAC^{e}(q_{1},\omega)$ and a $\mathcal{C}_{2}\in CAC^{e}(q_{2},\omega)$ into a $\mathcal{C}\in CAC^{e}(q_{1}q_{2},\omega)$ under certain conditions. One necessary condition for a $CAC^{e}$ of length $q_{1}q_{2}$ and weight $\omega$ being optimal is given. We also consider explicit construction of perfect $\mathcal{C}\in CAC^{e}(p,\omega)$ of odd prime $p$ and weight $\omega\geq3$. Finally, for positive integer $k$ and prime $p\equiv1 \ (\mbox{mod}\ 4k)$, we consider explicit construction of quasi-perfect $\mathcal{C}\in CAC^{e}(2p,4k+1)$.

\emph{$\mathbf{Index \ Terms-}$}Equi-difference conflict-avoiding code,\ Optimal code,\
Perfect code,\ Quasi-perfect code.

\section{Introduction}

The protocol sequence is one of the important topics of multiple-access communication system. In TDMA(Time Division Multiple Access), protocol sequence could be transformed from conflict-avoiding code and it have been investigated in \cite{LI,P,QLJ}. A conflict-avoiding code $(CAC)$ of length $n$ and weight $\omega$ is defined as a family $\mathcal{C}$ of $\omega$-subsets (called codewords) of $\mathbb{Z}_{n}=\mathbb{Z}/n\mathbb{Z}$ such that $\Delta(c_{1})\cap\Delta(c_{2})=\emptyset$  for any $c_{1},\ c_{2}\in\mathcal{C}$, $c_{1}\neq c_{2}$ where $\Delta(c)=\{j-i \ (\mbox{mod}\ n) \; | \; i,j\in c,i\neq j\}.$ Moreover, A $CAC\ \mathcal{C}$ is said to be equi-difference conflict-avoiding code $(CAC^{e})$ if every $c\in \mathcal{C}$ has the form $\{0,i,2i,\cdots,(\omega-1)i\}$ (see\cite{MSJ}). Let $CAC(n,\omega)$ denote the collection of all the $CACs$ of length $n$ and weight $\omega$. Similarly, the collection of all $(CAC^{e})s$ of length $n$ and weight $\omega$ is denoted by $CAC^{e}(n,\omega)$.
Let $M_{\omega}(n)=\mbox{max}\{|\mathcal{C}|: \mathcal{C}\in CAC(n,\omega)\}$ and we call a code $\mathcal{C}\in CAC(n,\omega)$ of size $M_{\omega}(n)$ optimal. Similarly, let $M_{\omega}^{e}(n)=\mbox{max}\{|\mathcal{C}|: \mathcal{C}\in CAC^{e}(n,\omega)\}$ and we call a code $\mathcal{C}\in CAC^{e}(n,\omega)$ of size $M_{\omega}^{e}(n)$ optimal. Some constructions for optimal $CACs$ of weight 3 and different even length can be found in \cite{FLM,JMJ,MFU,V} and the constructions for $\mathcal{C}\in CAC(n,3)$ of odd $n$ have been studied in \cite{K,KWC,WF,WCD,LMS}. As for $\omega>3$, various direct and recursive constructions of optimal $CACs$ for weight $\omega= 4,5$ have been obtained in \cite{MSJ}. Recently, Lin et al.\cite{LMJ} investigated sizes and constructions of optimal codes $\mathcal{C}\in CAC^{e}(n,4)$.

Let $\mathcal{C}\in CAC^{e}(n,\omega),$
$$\overline{\overline{x}}=\{0,x,2x,\cdots,(\omega-1)x\}$$  and
$$[\pm \omega]=\{-\omega,-\omega+1,\cdots,\omega-1,\omega\}.$$
If $\overline{\overline{x}}\in\mathcal{C}$, then
$$\Delta(\overline{\overline{x}})=\{ax \ (\mbox{mod}\ n)\; | \;a\in[\pm(\omega-1)]\backslash\{0\}\}.$$
Let $\mathcal{C}$ be a finite set consisting elements as $\overline{\overline{x}}(x\in\mathbb{Z}_{n})$. The set
$$\mathcal{C}\in CAC^{e}(n,\omega)$$
is equivalent to saying that for any $\overline{\overline{x}},\overline{\overline{y}}\in \mathcal{C}, \ a,b\in[\pm(\omega-1)]\ \mbox{with} \ (a,b)\neq(0,0),$
$$ax\equiv by \ (\mbox{mod}\ n)\Leftrightarrow a=b,\ x=y.$$

In addition, for a code $\mathcal{C}\in CAC^{e}(n,\omega)$, we have $2(\omega-1)|\mathcal{C}|+1\leq n,$ i.e.,
$$|\mathcal{C}|\leq\left\lfloor\frac{n}{2(\omega-1)}\right\rfloor.$$
If $|\mathcal{C}|=\frac{n}{2(\omega-1)}$, we call $\mathcal{C}$ a perfect code.
If $n\not\equiv1 \ (\mbox{mod}\ 2(\omega-1))$, then we call a code $\mathcal{C}$ of size $\left\lfloor\frac{n}{2(\omega-1)}\right\rfloor$ quasi-perfect. Obviously, perfect and quasi-perfect codes are optimal.

In this paper we focus on $CAC^{e}$s for general $\omega$. In Section 2 we show how to combine a $\mathcal{C}_{1}\in CAC^{e}(q_{1},\omega)$ and a $\mathcal{C}_{2}\in CAC^{e}(q_{2},\omega)$ into a $\mathcal{C}\in CAC^{e}(q_{1}q_{2},\omega)$ under certain conditions. One necessary condition for  a code $\mathcal{C}\in CAC^{e}(q_{1}q_{2},\omega)$ being optimal is given. Moreover, we can obtain many perfect or quasi-perfect $CAC^{e}$s through this method. In Section 3 we consider explicit construction of perfect $\mathcal{C}\in CAC^{e}(p,\omega)$ of odd prime $p$ and weight $\omega\geq3$; indeed, we get a necessary and sufficient condition to construct this  perfect $\mathcal{C}\in CAC^{e}(p,\omega)$. In Section 4, we construct quasi-perfect $\mathcal{C}\in CAC^{e}(2p,4k+1)$ for some primes $p$.

\section{Combining two $CAC^{e}$ into a new $CAC^{e}$}

The constructions of $CAC^{e}$ for some special length $n$ and weight $\omega$ have been obtained. Here, we consider how to combine two $CAC$s under certain conditions and the construction is inspired by Theorem 5 of \cite{KLY}.\\

$Theorem \ 1:$ Let $\mathcal{C}_{1}\in CAC^{e}(q_{1},\omega)$, $\mathcal{C}_{2}\in CAC^{e}(q_{2},\omega)$, and $\gcd(q_{2},(\omega-1)!)=1$. Let
$$\mathcal{C}_{1}\ltimes\mathcal{C}_{2}=\left\{\overline{\overline{x_{1}+q_{1}r}} \; | \; \overline{\overline{x_{1}}}\in\mathcal{C}_{1},r\in \mathbb{Z}_{q_{2}}\right\} \cup \left\{\overline{\overline{q_{1}x_{2}}} \; | \; \overline{\overline{x_{2}}}\in\mathcal{C}_{2}\right\}.$$
Then\\
\quad 1) \ $\mathcal{C}_{1}\ltimes\mathcal{C}_{2}\in CAC^{e}(q_{1}q_{2},\omega).$\\
\quad 2) \ $|\mathcal{C}_{1}\ltimes\mathcal{C}_{2}|=q_{2}|\mathcal{C}_{1}|+|\mathcal{C}_{2}|.$\\
\quad 3) \ $M_{\omega}^{e}(q_{1}q_{2})\geq q_{2}M_{\omega}^{e}(q_{1})+M_{\omega}^{e}(q_{2}).$

$Proof:$ 1) For $a,b\in [\pm(\omega-1)]$ with $(a,b)\neq(0,0)$, if
$$\overline{\overline{x}},\overline{\overline{y}}\in \left\{\overline{\overline{x_{1}+q_{1}r}} \; | \; \overline{\overline{x_{1}}}\in\mathcal{C}_{1},r\in \mathbb{Z}_{q_{2}}\right\},$$
then
$$\overline{\overline{x}}=\overline{\overline{x_{1}+q_{1}r_{1}}},\ \overline{\overline{y}}=\overline{\overline{y_{1}+q_{1}r_{2}}}$$
where $\overline{\overline{x_{1}}},\overline{\overline{y_{1}}}\in \mathcal{C}_{1}$ and $r_{1},r_{2}\in \mathbb{Z}_{q_{2}}.$\\
Indeed, if
$$a(x_{1}+q_{1}r_{1})\equiv b(y_{1}+q_{1}r_{2}) \ (\mbox{mod}\ q_{1}q_{2}),$$
then
$$ax_{1}\equiv by_{1} \ (\mbox{mod}\ q_{1})$$
which yields that $a=b$ and $x_{1}=y_{1}$. Since $\gcd(q_{2},(\omega-1)!)=1$, we have $r_{1}=r_{2}$
and so $x=y.$

Similarly, if
$$ \overline{\overline{x}},\ \overline{\overline{y}}\in \left\{\overline{\overline{q_{1}x_{2}}} \; | \; \overline{\overline{x_{2}}}\in\mathcal{C}_{2}\right\},$$
then
$$\overline{\overline{x}}=\overline{\overline{q_{1}x_{2}}}, \ \overline{\overline{y}}=\overline{\overline{q_{1}y_{2}}}$$
where $\overline{\overline{x_{2}}},\ \overline{\overline{y_{2}}}\in \mathcal{C}_{2}.$\\
Indeed, if
$$aq_{1}x_{2}\equiv bq_{1}y_{2} \ (\mbox{mod}\ q_{1}q_{2}),$$
then
$$ax_{2}\equiv by_{2} \ (\mbox{mod}\ q_{2}).$$
Since $\overline{\overline{x_{2}}},\overline{\overline{y_{2}}}\in \mathcal{C}_{2},$ we have $a=b$ and $x_{2}=y_{2}$. Hence $x=y.$

Finally, if
$$\overline{\overline{x}}\in \left\{\overline{\overline{x_{1}+q_{1}r}} \; | \; \overline{\overline{x_{1}}}\in\mathcal{C}_{1},r\in \mathbb{Z}_{q_{2}}\right\} \ \mbox{and} \ \overline{\overline{y}}\in \left\{\overline{\overline{q_{1}x_{2}}} \; | \; \overline{\overline{x_{2}}}\in\mathcal{C}_{2}\right\},$$
then
$$\overline{\overline{x}}=\overline{\overline{x_{1}+q_{1}r_{1}}}, \ \overline{\overline{y}}=\overline{\overline{q_{1}y_{2}}}$$
where $\overline{\overline{x_{1}}}\in \mathcal{C}_{1}$, $\overline{\overline{y_{2}}}\in \mathcal{C}_{2}$ and $r_{1}\in\mathbb{Z}_{q_{2}}.$\\
Indeed, if
$$a(x_{1}+q_{1}r_{1})\equiv bq_{1}y_{2} \ (\mbox{mod}\ q_{1}q_{2}),$$
then
$$ax_{1}\equiv 0 \ (\mbox{mod}\ q_{1}).$$
We have $a=0$ since $\overline{\overline{x_{1}}}\in \mathcal{C}_{1}$. Hence
$$0\equiv by_{2} \ (\mbox{mod}\ q_{2}).$$
This implies that $b=0$ which contradicts to the assumption that $(a,b)\neq(0,0).$

Part 2) and Part 3) can be easily derived.\hfill$\blacksquare$\\

$Lemma \ 1:$ If $\mathcal{C}\in CAC^{e}(q_{1}q_{2},\omega),$ then
$$\frac{1}{q_{2}}\left(\mathcal{C}\cap \left\{\overline{\overline{q_{2}z}} \; | \;z\in\mathbb{Z}_{q_{1}}\right\}\right)\in CAC^{e}(q_{1},\omega).$$

$Proof:$ For $\overline{\overline{x}}, \ \overline{\overline{y}}\in\frac{1}{q_{2}}\left(\mathcal{C}\cap \left\{\overline{\overline{q_{2}z}} \; | \; z\in\mathbb{Z}_{q_{1}}\right\}\right)\in CAC^{e}(q_{1},\omega),$ one has
$$\overline{\overline{q_{2}x}}, \ \overline{\overline{q_{2}y}}\in \mathcal{C}.$$
Suppose that $ax\equiv by \ (\mbox{mod}\ q_{1})$ for $a,b\in [\pm(\omega-1)]$ with $(a,b)\neq(0,0).$ Then
$$aq_{2}x\equiv bq_{2}y \ (\mbox{mod}\ q_{1}q_{2})$$
which implies $a=b$ and $x=y.$\hfill$\blacksquare$\\

$Remark \ 1.$ Similarly, we have $\frac{1}{q_{1}}\left(\mathcal{C}\cap \left\{\overline{\overline{q_{1}z}} \; | \; z\in \mathbb{Z}_{q_{2}}\right\}\right)\in CAC^{e}(q_{2},\omega).$ More generally, for $\mathcal{C}\in CAC^{e}(n,\omega)$ and $q_{i}$ being positive factor of $n$, let $\mathcal{C}_{i}^{e}=\frac{q_{i}}{n}\left(\mathcal{C}\cap
\left\{\overline{\overline{\frac{n}{q_{i}}z}} \; | \; z\in\mathbb{Z}_{q_{i}}\right\}\right).$ Then $\mathcal{C}_{i}^{e}\in CAC^{e}(q_{i},\omega).$\\

$Theorem \ 2:$ Let $\mathcal{C}\in CAC^{e}(q_{1}q_{2},\omega).$ If $|\mathcal{C}_{i}^{e}|\neq M_{\omega}^{e}(q_{i})$ and $\gcd(\frac{q_{1}q_{2}}{q_{i}},(\omega-1)!)=1$ for some $i\in\{1,2\}$, then $|\mathcal{C}|\neq M_{\omega}^{e}(q_{1}q_{2}).$

$Proof:$ Without loss of generality, let $|\mathcal{C}_{1}^{e}|\neq M_{\omega}^{e}(q_{1})$ and $\mathcal{C}_{1}^{M}\in CAC^{e}(q_{1},\omega)$ with $|\mathcal{C}_{1}^{M}|= M_{k}^{e}(q_{1}).$ In the following we will show $\mathcal{C}'=\left(\mathcal{C}\backslash \{q_{2}c \; | \; c\in\mathcal{C}_{1}^{e}\}\right)\cup \{q_{2}c \; | \; c\in\mathcal{C}_{1}^{M}\}\in CAC^{e}(q_{1}q_{2},\omega)$ and $|\mathcal{C}'|>|\mathcal{C}|.$

 Let $\overline{\overline{x}}\ \overline{\overline{y}}\in \mathcal{C}'$, $a,b\in [\pm(\omega-1)]$ with $(a,b)\neq(0,0).$ Obviously, if
 $$\overline{\overline{x}},\overline{\overline{y}}\in\mathcal{C}\backslash \{q_{2}c \; | \; c\in\mathcal{C}_{1}^{e}\} \ \mbox{and} \ ax \equiv by \ (\mbox{mod}\ q_{1}q_{2}),$$
 then $a=b$ and $x=y.$\\
 If
 $$ \overline{\overline{x}},\overline{\overline{y}}\in \{q_{2}c|c\in\mathcal{C}_{1}^{M}\},$$
 then
 $$\overline{\overline{x}}=\overline{\overline{q_{2}x_{M}}} \ \mbox{and} \ \overline{\overline{y}}=\overline{\overline{q_{2}y_{M}}}$$
 where $\overline{\overline{x_{M}}}, \ \overline{\overline{y_{M}}}\in\mathcal{C}_{1}^{M}.$
 Indeed, if
 $$ax \equiv by \ (\mbox{mod}\ q_{1}q_{2}), \ \mbox{i.e.},$$
 $$aq_{2}x_{M}\equiv bq_{2}y_{M} \ (\mbox{mod}\ q_{1}q_{2}),$$
 then
 $$ax_{M}\equiv by_{M} \ (\mbox{mod}\ q_{1})$$
 which yields $a=b$, $x_{M}=y_{M}$ and so $x=y.$ \\
 For
 $$\overline{\overline{x}}\in\mathcal{C}\backslash \{q_{2}c \; | \; c\in\mathcal{C}_{1}^{e}\} \  \mbox{and} \ \overline{\overline{y}}\in \{q_{2}c \; | \; c\in\mathcal{C}_{1}^{M}\},$$
 we have $\overline{\overline{x}}=\overline{\overline{\alpha+q_{2}r_{1}}}$ with $r_{1}\in \mathbb{Z}_{q_{1}}$ and $1 \leq\alpha<q_{2}$ by Remark 1.
 If $ax\equiv by \ (\mbox{mod}\ q_{1}q_{2}),$ i.e.,
 $$a(\alpha+q_{2}r_{1})\equiv bq_{2}y_{M} \ (\mbox{mod}\ q_{1}q_{2})$$
 with
 $$\overline{\overline{y_{M}}}\in\mathcal{C}_{1}^{M},1 \leq\alpha<q_{2},\ \mbox{and} \ r_{1}\in \mathbb{Z}_{q_{1}}$$
 which implies
 $$a\alpha\equiv 0 \ (\mbox{mod}\ q_{2}).$$
 We have $\alpha=0$ since $\gcd(\frac{q_{1}q_{2}}{q_{i}},(\omega-1)!)=1$ which contradicts to the assumption that $1\leq\alpha<q_{2}.$

 Hence, $\mathcal{C}'\in CAC^{e}(q_{1}q_{2},k)$ and $|\mathcal{C}'|=|\mathcal{C}|-|\mathcal{C}|+M_{k}^{e}(q_{1})>|\mathcal{C}|.$\hfill$\blacksquare$\\

$Corollary \ 1:$ Let $\gcd(n,(\omega-1)!)=1$ and $\mathcal{C}\in CAC^{e}(n,\omega)$ with $|\mathcal{C}|= M_{\omega}^{e}(n).$ Then $|\mathcal{C}_{i}^{e}|= M_{\omega}^{e}(q_{i})$ for $q_{i}$ being any positive factor of $n$.\\

$Corollary \ 2:$ Let $n=q_{1}q_{2}\cdots q_{t}$ and $\mathcal{C}\in CAC^{e}(n,\omega)$ be perfect. If $\gcd(\frac{n}{q_{i}},(\omega-1)!)=1$ for some $i$ and there exists perfect  $CAC^{e}$s in $\mathbb{Z}_{q_{i}}$, then $\mathcal{C}_{i}^{e}$ is perfect. Conversely, let $n=q_{1}q_{2}\cdots q_{t}$. If $\gcd(n,(\omega-1)!)=1$ and $\mathcal{C}_{i}^{e}\in CAC^{e}(q_{i},\omega)(1\leq i\leq t)$ is perfect, then $\ltimes_{i=1}^{t}\mathcal{C}_{i}\in CAC^{e}(n,\omega)$ is perfect.\\

$Corollary \ 3:$ Let $n=q_{1}q_{2}\cdots q_{t}$, $\gcd(\frac{n}{q_{i}},(\omega-1)!)=1$, and $\mathcal{C}\in CAC^{e}(n,\omega)$ be quasi-perfect. If there exists perfect or quasi-perfect $CAC^{e}$s in $\mathbb{Z}_{q_{i}}$, then $\mathcal{C}_{i}^{e}$ is perfect or quasi-perfect. Conversely,  there is a quasi-perfect $\mathcal{C}_{j}^{e}\in CAC^{e}(q_{j},\omega)$ for some $j$ and each $\mathcal{C}_{i}^{e}\in CAC^{e}(q_{i},\omega)$ with $i\neq j$ is perfect, then $(\ltimes_{i\neq j}\mathcal{C}_{i}^{e})\ltimes\mathcal{C}_{j}^{e}\in CAC^{e}(q_{1}q_{2}\cdots q_{t},\omega)$ is quasi-perfect.\\

$Lemma \ 2:$ For any $n=(2\omega-1)q_{1}$ with $\gcd((\omega-1)!,q_{1})=1,$ we have $M_{\omega}^{e}(n)\geq q_{1}+M_{\omega}^{e}(q_{1}).$ Moreover, a perfect or quasi-perfect $CAC^{e}(n,\omega)$ exists if a perfect or quasi-perfect $CAC^{e}(q_{1},\omega)$ exists. In particular, \\
\quad 1) if $\omega-1<q_{1}<2(\omega-1)$, then $\left\{\overline{\overline{(2\omega-1)r_{1}+1}} \; | \; r_{1}\in\mathbb{Z}_{q_{1}}\right\}\in CAC^{e}(n,\omega)$ is quasi-perfect and it has $q_{1}$ codewords; \\
\quad 2) if $q_{1}=2\omega-1$, then $\left\{\overline{\overline{1}}\right\}\ltimes\left\{\overline{\overline{1}}\right\}\in CAC^{e}(n,\omega)$ is perfect and it has $2\omega$ codewords; \\
\quad 3) if $2\omega-1<q_{1}<4(\omega-1)$, then $\left\{\overline{\overline{1}}\right\}\ltimes\left\{\overline{\overline{1}}\right\}\in CAC^{e}(n,\omega)$ is quasi-perfect and it has $q_{1}+1$ codewords.

$Proof:$ Obviously, $\left\{\overline{\overline{1}}\right\}\in CAC^{e}(2\omega-1,\omega)$ is perfect. If there exists a perfect or quasi-perfect code $\mathcal{C}_{1}\in CAC^{e}(q_{1},\omega)$, then $\left\{\overline{\overline{1}}\right\}\ltimes\mathcal{C}_{1}\in CAC^{e}(n,\omega)$ is perfect or quasi-perfect by Corollary 2 and Corollary 3. Precisely, \\
1) if $\omega-1<q_{1}<2(\omega-1)$, then $\emptyset\in CAC^{e}(q_{1},\omega)$ is quasi-perfect.
Hence $$\left\{\overline{\overline{1}}\right\}\ltimes\emptyset=\{\overline{\overline{(2\omega-1)r_{1}+1}} \; | \; r_{1}\in\mathbb{Z}_{q_{1}}\}\in CAC^{e}(n,\omega)$$
is quasi-perfect and $|\left\{\overline{\overline{(2\omega-1)r_{1}+1}} \; | \; r_{1}\in\mathbb{Z}_{q_{1}}\right\}|=q_{1}$.

The proofs of 2) and 3) are  similar to that of 1).\hfill$\blacksquare$\\

$Example \ 1.$ Let $\omega=3$ and $q=65=5\times 13$. Then $$\mathcal{C}=\left\{\overline{\overline{1}}, \overline{\overline{3}}, \overline{\overline{4}}, \overline{\overline{9}} , \overline{\overline{10}}, \overline{\overline{12}}, \overline{\overline{13}}, \overline{\overline{14}}, \overline{\overline{16}}, \overline{\overline{17}}, \overline{\overline{22}}, \overline{\overline{23}}, \overline{\overline{27}}, \overline{\overline{29}}, \overline{\overline{30}}\right\}\in CAC^{e}(65,3)$$
 and $$\mathcal{C}_{2}^{e}=\left\{\overline{\overline{2}},\overline{\overline{6}}\right\}\in CAC^{e}(13,3).$$
 However, $\left\{\overline{\overline{1}},\overline{\overline{3}},\overline{\overline{4}}\right\}\in CAC^{e}(13,3)$ is  perfect. Thus, $|\mathcal{C}|\neq M_{3}^{e}(65)$ by Theorem 2 and $$\left\{\overline{\overline{1}},\overline{\overline{3}},\overline{\overline{4}}\right\}\ltimes\left\{\overline{\overline{1}}\right\}=\left\{\overline{\overline{1}}, \overline{\overline{3}}, \overline{\overline{4}}, \overline{\overline{13}}, \overline{\overline{14}}, \overline{\overline{16}}, \overline{\overline{17}}, \overline{\overline{27}}, \overline{\overline{29}}, \overline{\overline{30}}, \overline{\overline{40}}, \overline{\overline{42}}, \overline{\overline{43}}, \overline{\overline{53}}, \overline{\overline{55}}, \overline{\overline{56}}\right\}\in CAC^{e}(65,3)$$
  is perfect.\\

$Example \ 2.$ Let $\omega=6$ and $q=121=11^{2}$. Then $(11,5!)=1$ and $\left\{\overline{\overline{1}}\right\}\in CAC^{e}(11,6)$ is perfect. Thus, $\mathcal{C}=\left\{\overline{\overline{1}}\right\}\ltimes\left\{\overline{\overline{1}}\right\}=\left\{\overline{\overline{1}},\overline{\overline{5}},\overline{\overline{6}},\overline{\overline{11}},\overline{\overline{16}},
\overline{\overline{21}},\overline{\overline{26}},\overline{\overline{31}},\overline{\overline{36}},\overline{\overline{41}},
\overline{\overline{46}},\overline{\overline{51}}\right\}$ is a perfect code in $CAC^{e}(121,6)$ by Corollary  2.\\

\section{Perfect $CAC^{e}$ of prime length}

Let $\lambda=\omega-1$. We know that a perfect $CAC^{e}$ of length $n$ and weight $\omega$  is a code $\mathcal{C}\in CAC^{e}(n,\omega)$ with $2\lambda|\mathcal{C}|+1=n.$ For length $n=2$, there is no perfect code $\mathcal{C}\in CAC^{e}(2,\omega)$. For odd prime $p$, $\left\{\{0,i\} \; | \; i=1,2,\cdots,\frac{p-1}{2}\right\}$ is a perfect $CAC^{e}(p,2)$. Let $g$ be a primitive root modulo $p$ and
$$\mu=\gcd\left(ind_{g}(-1),ind_{g}(2),ind_{g}(3),\cdots,ind_{g}(\lambda)\right)$$
where $ind_{g}(a)$ is the index of $a$ relative to the base $g$, i.e.,
$$a\equiv g^{ind_{g}(a)} \ (\mbox{mod}\ p) \ \mbox{and} \ 0\leq ind_{g}(a)<p-1.$$
It is clear that the set
$$H=\{g^{i\mu} \ (\mbox{mod}\ p)|i\geq0\}$$
is the multiplicative subgroup of $\mathbb{Z}_{p}^{*}$ generated by the integers $-1,2,3,\cdots,\lambda$. Moreover, if a perfect code $\mathcal{C}\in CAC^{e}(p,\omega)$ exists with $\omega\geq3$, then $p\equiv1 \ (\mbox{mod}\ 2\mu\lambda)$ by  Theorem 1 of \cite{KLY}.
For an odd prime $p$ and $\omega\geq3$, we will consider explicit construction of perfect code $\mathcal{C}\in CAC^{e}(p,\omega)$.\\

$Theorem \ 3:$ Let $\omega\geq3$ and $p$ be a prime such that $p\equiv1 \ (\mbox{mod}\ 2\mu\lambda)$. Let $g$ be a primitive root modulo $p$. We have
$$\qquad\qquad\qquad\qquad\qquad\quad \mathcal{C}=\left\{\overline{\overline{g^{\mu\lambda i+j} \ (\mbox{mod}\ p)}} \; | \; i\in [0,\frac{p-1}{2\mu\lambda}-1],j\in[0,\mu-1]\right\}\qquad\qquad\qquad\qquad\qquad\qquad (1)$$
is a perfect $CAC^{e}$ of length $n$ and weight $\omega$  if and only if
$$\quad\quad\qquad\qquad\qquad\qquad\qquad\qquad\left\{\frac{ind_{g}(k)}{\mu} \ (\mbox{mod}\ \lambda) \; | \; k\in[1,\lambda]\right\}=[0,\lambda-1].\qquad\qquad\qquad\qquad\qquad\qquad\qquad(2)$$

$Proof:$ Suppose $\left\{\frac{ind_{g}(k)}{\mu} \ (\mbox{mod}\ \lambda) \; | \; k\in[1,\lambda]\right\}\neq[0,\lambda-1],$ i.e., there are $k_{1}\neq k_{2}\in[1,\lambda]$ such that
$$\frac{ind_{g}(k_{1})}{\mu} \ (\mbox{mod}\ \lambda)=\frac{ind_{g}(k_{2})}{\mu} \ (\mbox{mod}\ \lambda).$$
Since $k_{1}\neq k_{2}\in[1,\lambda]$, there exists integer $N\neq0,\frac{p-1}{2}$ such that
$$ind_{g}(k_{1})=ind_{g}(k_{2})+N\mu\lambda.$$
We define $S(N)$ as follows
$$S(N)=
\left\{
\begin{array}{ll}
1 ,&  N\mu\lambda \ (\mbox{mod}\ p-1)<\frac{p-1}{2},\\
-1 ,&  N\mu\lambda \ (\mbox{mod}\ p-1)\geq\frac{p-1}{2}.
\end{array}
\right.$$
Thus $$ind_{g}(k_{1})\equiv ind_{g}(S(N)k_{2})+N'\mu\lambda \ (\mbox{mod}\ p-1),$$
with $N'\in[1,\frac{p-1}{2\mu\lambda}-1]$ since $N\neq0,\frac{p-1}{2}$.
Hence $k_{1}g^{0}\equiv S(N)k_{2}g^{N'\mu\lambda} \ (\mbox{mod}\ p)$ with $k_{1},S(N)k_{2}\in [\pm\lambda]\setminus\{0\}$ and $\overline{\overline{g^{0}}}\neq\overline{\overline{g^{N'\mu\lambda} \ (\mbox{mod}\ p)}}\in \mathcal{C}$  which contradicts to the assumption that $\mathcal{C}\in CAC^{e}(p,\omega)$ is perfect.

It's easy to see that $|\mathcal{C}|=\frac{p-1}{2\lambda}$ and in the following we will show $\mathcal{C}\in CAC^{e}(p,\omega)$.
Clearly,
$$ag^{\mu\lambda i+j}\not\equiv0 \ (\mbox{mod}\ p)\ \mbox{for \ any} \ a\in[1,\lambda].$$
For $a,b\in [\pm\lambda]\setminus\{0\}$, we have $a,b\in H$, i.e.,
$$a=g^{\mu r_{1}} \ (\mbox{mod}\ p),b=g^{\mu r_{2}} \ (\mbox{mod}\ p).$$
For $\overline{\overline{g^{i_{1}\mu\lambda+j_{1}}}},\ \overline{\overline{g^{i_{2}\mu\lambda+j_{2}}}}\in \mathcal{C}$, if
$$g^{\mu r_{1}}g^{i_{1}\mu\lambda+j_{1}}\equiv g^{\mu r_{2}}g^{i_{2}\mu\lambda+j_{2}} \ (\mbox{mod}\ p),$$
then
$$\mu r_{1}+i_{1}\mu\lambda+j_{1}\equiv \mu r_{2}+i_{2}\mu\lambda+j_{2} \ (\mbox{mod}\ p-1).$$
Modulo $\mu$ we get
$$j_{1}\equiv j_{2} \ (\mbox{mod}\ \mu)$$ which implies $j_{1}=j_{2}$ since $j_{1},j_{2}\in [0,\mu-1].$
Therefore,
$$i_{1}\lambda+r_{1}\equiv i_{2}\lambda+r_{2} \ \left(\mbox{mod}\ \frac{p-1}{\mu}\right)$$
and so $$r_{1}\equiv r_{2} \ (\mbox{mod}\ \lambda).$$
Combining with (2), this implies that $a=b$ or $a=-b.$\\
If $a=b,$ then $r_{1}=r_{2}$ which implies that
$$i_{1}\equiv i_{2} \ \left(\mbox{mod}\ \frac{p-1}{\mu\lambda}\right) \ \mbox{and so} \ i_{1}=i_{2}.$$
Hence, $g^{i_{1}\mu\lambda+j_{1}}\equiv g^{i_{2}\mu\lambda+j_{2}} \ (\mbox{mod}\ p).$\\
If $a=-b,$ then
$$i_{1}\neq i_{2} \ \mbox{and} \ i_{1}\mu\lambda \equiv i_{2}\mu\lambda+\frac{p-1}{2} \ (\mbox{mod}\ p-1).$$
Modulo $\frac{p-1}{2}$ we obtain
$$i_{1}\mu\lambda\equiv i_{2}\mu\lambda \ \left(\mbox{mod}\ \frac{p-1}{2}\right)$$
and so $i_{1}\equiv i_{2} \ \left(\mbox{mod}\ \frac{p-1}{2\mu\lambda}\right),$ i.e.,
$$i_{1}=i_{2}$$
which is a contradiction.\hfill$\blacksquare$\\

$Remark \ 2.$ For $v$ being any positive factor of $u$ and $\gcd(\mu/v,\lambda)=1$, we have $\left\{\frac{ind_{g}(k)}{v} \ (\mbox{mod}\ \lambda) \; | \; k\in[1,\lambda]\right\}=[0,\lambda-1]$ when $\left\{\frac{ind_{g}(k)}{\mu} \ (\mbox{mod}\ \lambda) \; | \; k\in[1,\lambda]\right\}=[0,\lambda-1].$ Thus, the formula (1) of Theorem 3 can be modified as
$$\mathcal{C}=\left\{\overline{\overline{g^{v\lambda i+j} \ (\mbox{mod}\ p)}} \; | \; i\in [0,\frac{p-1}{2v\lambda}-1],j\in[0,v-1]\right\}$$ by the proof of Theorem 3.\\

$Example \ 3.$ Let $\omega=5$ and $p=97$. Then choose $g=5$ being a primitive root,
$$\mu=\gcd\left(ind_{g}(-1),ind_{g}(2),ind_{g}(3),ind_{g}(4)\right)=\gcd(48,34,70,68)=2,$$
and
$$\left\{\frac{ind_{g}(k)}{2} \ (\mbox{mod}\ \lambda) \; | \; k\in[1,\lambda]\right\}=\left\{0,\frac{34}{2} \ (\mbox{mod}\ 4),\frac{70}{2} \ (\mbox{mod}\ 4) ,\frac{68}{2} \ (\mbox{mod}\ 4)\right\}=\{0,1,3,2\}=[0,3].$$
Hence
$$\mathcal{C}=\left\{\overline{\overline{g^{\mu\lambda i+j} \ (\mbox{mod}\ p)}} \; | \; i\in [0,\frac{p-1}{2\mu\lambda}-1],j\in[0,\mu-1]\right\}=\left\{\overline{\overline{1}}, \overline{\overline{5}}, \overline{\overline{6}}, \overline{\overline{30}}, \overline{\overline{36}}, \overline{\overline{83}}, \overline{\overline{22}}, \overline{\overline{13}}, \overline{\overline{35}}, \overline{\overline{78}}, \overline{\overline{16}}, \overline{\overline{80}}\right\}$$
is a perfect $CAC^{e}$ of length 97 and weight 5.

In Table 1  we give some examples of the first primes that satisfy the conditions of Theorem 3 for $\omega\in[5,12].$

\begin{table}
\caption{Perfect\ code\ $\mathcal{C}\in CAC^{e}(p,\omega)$}
\centering
\begin{sideways}
   \begin{tabular}{|c|c|cccccccccccc|}
   \hline
   $\omega$ & \multicolumn{13}{|c|}{$\mathcal{C}=\{\overline{\overline{g^{\mu\lambda i+j} \ (\mbox{mod}\ p)}} \; | \; i\in [0,\frac{p-1}{2\mu\lambda}-1],j\in[0,\mu-1]\}$}\\
   \hline

   \multirow{3}{*}{5}
& $p$ & 97 & 409 & 1201 & 1873 & 2161 & 2617 & 3433 & 3457 & 3529 & 5233 & 5641 & 6577 \\[-1mm]
\cline{2-14}
& $g$ & 5 & 21 & 11 & 10 & 23 & 5 & 5 & 7 & 17 & 10 & 14 & 5 \\[-1mm]
\cline{2-14}
& $\mu$ & 2 & 2 & 4 & 2 & 2 & 2 & 2 & 2 & 4 & 4 &2 & 2 \\[-1mm]
\hline
\multirow{3}{*}{6}
& $p$ & 11 & 421 & 701 & 2311 & 2861 & 3187 & 3491 & 3931 & 4621 & 5531 & 6121 & 7621  \\[-1mm]
\cline{2-14}
& $g$ & 2 & 2 & 2 & 3 & 2 & 7 & 2 & 2 & 2 & 10 & 7 & 2 \\[-1mm]
\cline{2-14}
& $\mu$ & 1 & 1 & 1 & 1 & 1 & 1 & 1 & 1 & 1 & 1 &2 & 1 \\[-1mm]
\hline
\multirow{3}{*}{7}
& $p$ & 13 & 769 & 1249 & 2521 & 3049 & 5881 & 7477 & 7933 & 8293 & 9769 & 10837 & 12049  \\[-1mm]
\cline{2-14}
& $g$ & 2 & 11 & 7 & 17 & 11 & 31 & 2 & 2 & 2 & 13 & 2 & 13 \\[-1mm]
\cline{2-14}
& $\mu$ & 1 & 2 & 2 & 2 & 2 & 2 & 1 & 1 & 1 & 2 & 1 & 2 \\[-1mm]
\hline
\multirow{3}{*}{8}
& $p$ & 659 & 1429 & 2087 & 3557 & 4633 & 9689 & 12391 & 17431 & 20749 & 21001 & 21911 & 28211  \\[-1mm]
\cline{2-14}
& $g$ & 2 & 6 & 5 & 2 & 3 & 3 & 26 & 3 & 2 & 11 & 13 & 2\\[-1mm]
\cline{2-14}
& $\mu$ & 1 & 1 & 1 & 1 & 1 & 1 & 1 & 1 & 1 & 2 & 1 & 1 \\[-1mm]
\hline
\multirow{3}{*}{9}
& $p$ & 17 & 3617 & 6257 & 15377 & 21377 & 22193 & 42257 & 48049 & 61441 & 77153 & 78497 & 81233  \\[-1mm]
\cline{2-14}
& $g$ & 3 & 3 & 3 & 3 & 3 & 3 & 3 & 17 & 17 & 3 & 3 & 3 \\[-1mm]
\cline{2-14}
& $\mu$ & 1 & 1 & 1 & 1 & 1 & 1 & 1 & 2 & 2 & 1 & 1 & 1  \\[-1mm]
\hline
\multirow{3}{*}{10}
& $p$ & 19 & 27127 & 30241 & 30781 & 47017 & 59473 & 86599 & 162109 & 243829 & 268003 & 271729 & 276373  \\[-1mm]
\cline{2-14}
& $g$ & 2 & 3 & 11 & 2 & 7 & 10 & 3 & 2 & 2 & 2 & 11 & 2 \\[-1mm]
\cline{2-14}
& $\mu$ & 1 & 1 & 2 & 1 & 1 & 1 & 1 & 1 & 1 & 1 & 1 & 1 \\[-1mm]
\hline
\multirow{3}{*}{11}
& $p$ & 3181 & 211741 & 214021 & 274861 & 289141 & 298861 & 348421 & 447901 & 531901 & 619261 & 661741 & 691381  \\[-1mm]
\cline{2-14}
& $g$ & 7 & 2 & 2 & 2 & 2 & 7 & 2 & 2 & 6 & 6 & 2 & 2 \\[-1mm]
\cline{2-14}
& $\mu$ & 1 & 1 & 1 & 1 & 1 & 1 & 1 & 1 & 1 & 1 & 1 & 1 \\[-1mm]
\hline
\multirow{3}{*}{12}
& $p$ & 23 & 56431 & 78541 & 218989 & 591559 & 631357 & 1059257 & 1133551 & 1588423 & 1768229 & 1797379 & 2079419 \\[-1mm]
\cline{2-14}
& $g$ & 5 & 3 & 2 & 14 & 6 & 2 & 3 & 3 & 5 & 2 & 2 & 2 \\[-1mm]
\cline{2-14}
& $\mu$ & 1 & 1 & 1 & 1 & 1 & 1 & 1 & 1 & 1 & 1 & 1 & 1 \\
\hline
   \end{tabular}

\end{sideways}
\end{table}

\section{On quasi-perfect code $\mathcal{C}\in CAC^{e}(2p,4k+1)$}

Since $n\equiv1 \ (\mbox{mod}\ 2(\omega-1))$ is a necessary condition for the existence of a perfect code $\mathcal{C}\in CAC^{e}(n,\omega)$, perfect $CAC^{e}$ can not exist if $\gcd(\omega-1,n)>1.$ However, quasi-perfect $CAC^{e}$ may well exist. In this section, for $\gcd(\omega-1,n)=2,$ we consider explicit construction of quasi-perfect $\mathcal{C}\in CAC^{e}(2p,4k+1)$ of positive integer $k$ and prime $p\equiv1 \ (\mbox{mod}\ 4k)$.\\

$Lemma \ 3:$ Let $p\equiv1 \ (\mbox{mod}\ 4k)$ be a prime where $k$ is a positive integer and let $g_{1},\ g_{2}$ be two primitive elements of $\mathbb{Z}_{p}$. If $$\{ind_{g_{1}}(i) \ (\mbox{mod}\ 2k)|i=1,3,\cdots,4k-1\}=\{ind_{g_{1}}(i) \ (\mbox{mod}\ 2k) \; | \; i=2,4,\cdots,4k\}=[0,2k-1],$$ then
$$\{ind_{g_{2}}(i) \ (\mbox{mod}\ 2k) \; | \; i=1,3,\cdots,4k-1\}=\{ind_{g_{2}}(i) \ (\mbox{mod}\ 2k)|i=2,4,\cdots,4k\}=[0,2k-1].$$

$Proof:$ Firstly, $g_{2}=g_{1}^{s}$ for some $s$ with $(s,p-1)=1$ and so
$$ind_{g_{1}}(i)\equiv ind_{g_{2}}(i)s \ (\mbox{mod}\ p-1).$$
Modulo $2k$ we get
$$ind_{g_{1}}(i)\equiv ind_{g_{2}}(i)s \ (\mbox{mod}\ 2k).$$
Also, we note that $ind_{g_{2}}(i)$ will run through $\mathbb{Z}_{2k}$ when $ind_{g_{1}}(i)$ does since $(s,p-1)=1$, i.e.,
$$\{ind_{g_{2}}(i) \ (\mbox{mod}\ 2k)\; | \; i=1,3,\cdots,4k-1\}=\{ind_{g_{2}}(i) \ (\mbox{mod}\ 2k) \; | \; i=2,4,\cdots,4k\}=[0,2k-1].$$
\hfill$\blacksquare$\\

$Theorem \ 4:$ Let $p\equiv1 \ (\mbox{mod}\ 4k)$ be a prime where $k$ is a positive integer and let $g_{1}$ be a primitive element of $\mathbb{Z}_{p}$. If $$\{ind_{g_{1}}(i) \ (\mbox{mod}\ 2k) \; | \; i=1,3,\cdots,4k-1\}=\{ind_{g_{1}}(i) \ (\mbox{mod}\ 2k) \; | \; i=2,4,\cdots,4k\}=[0,2k-1],$$
then for $g$ being an odd primitive element of $\mathbb{Z}_{p}$, the set
$$\mathcal{C}=\left\{\overline{\overline{g^{2kj}}} \ (\mbox{mod}\ 2p) \; | \; 0\leq j<(p-1)/4k\right\}$$
is a quasi-perfect code in $CAC^{e}(2p,4k+1)$.

$Proof:$ Suppose
$$ag^{2kj_{1}}\equiv bg^{2kj_{2}} \ (\mbox{mod}\ 2p)$$
for some $a,\ b\in[\pm4k]$, $(a,b)\neq(0,0)$ and
$$0\leq j_{1},\ j_{2}<(p-1)/4k.$$

If $a=0,$ i.e., $bg^{2kj}\equiv0 \ (\mbox{mod}\ 2p)$, then $b\equiv0 \ (\mbox{mod}\ p)$ and so $b=0.$

Otherwise both $a$ and $b$ are nonzero. Firstly
$$a\equiv b \ (\mbox{mod}\ 2).$$
Secondly
$$ind_{g}(a)+2kj_{1}\equiv ind_{g}(b)+2kj_{2} \ (\mbox{mod}\ p-1)$$
which yields
$$ind_{g}(a)\equiv ind_{g}(b) \ (\mbox{mod}\ 2k).$$
We note that
$$ind_{g}(-a)\equiv ind_{g}(a)+(p-1)/2\equiv ind_{g}(a) \ (\mbox{mod}\ 2k).$$
Thus by lemma 3 we have $a=\pm b.$

If $a=b$, then
$$2kj_{1}\equiv2kj_{2} \ (\mbox{mod}\ p-1).$$
and so
$$j_{1}\equiv j_{2} \ (\mbox{mod}\ (p-1)/2k), \  \mbox{i.e.,} \ j_{1}=j_{2}.$$
If $a=-b,$ then $j_{1}\neq j_{2}$ and
$$2kj_{1}\equiv 2kj_{2}+(p-1)/2 \ (\mbox{mod}\ p-1).$$
Modulo $\frac{p-1}{2}$ we get
$$2kj_{1}\equiv 2kj_{2} \ (\mbox{mod}\ (p-1)/2)$$
and so
$$j_{1}\equiv j_{2} \ (\mbox{mod}\ (p-1)/4k),\  \mbox{i.e.,} \ j_{1}=j_{2}$$
which is a contradiction.

Thus $\mathcal{C}=\left\{\overline{\overline{g^{2kj} \ (\mbox{mod}\ 2p)}} \; | \; 0\leq j<(p-1)/4k\right\}$
is a quasi-perfect code in $CAC^{e}(2p,4k+1)$ and $|\mathcal{C}|=\left\lfloor\frac{2p-1}{8k}\right\rfloor=\frac{p-1}{4k}$.\hfill$\blacksquare$\\

$Example \ 3.$ Let $k=1$, $p=29$. Choose $g=3$. Then
$$\{ind_{g}(i) \; | \;  i \ \in [1,4]\}=\{0, 17, 1, 6\}$$
and so
$$\{ind_{g}(i) \ (\mbox{mod}\ 2) \; | \; i=1,3\}=\{0, 1\}=[0,1],$$
$$\{ind_{g}(i) \ (\mbox{mod}\ 2) \; | \; i=2,4\}=\{1, 0\}=[0,1].$$
Since $g^{2}\equiv 9 \ (\mbox{mod}\ 58),$ we get the following  quasi-perfect code in $CAC^{e}(58,5)$:
$$\mathcal{C}=\left\{\overline{\overline{9^{j} \ (\mbox{mod}\ 58)}} \; | \; 0\leq j<7\right\}=\left\{\overline{\overline{1}},\overline{\overline{9}},\overline{\overline{23}},\overline{\overline{33}},\overline{\overline{7}},\overline{\overline{5}},
\overline{\overline{45}}\right\}.$$
We note that we can replace any elements $\overline{\overline{c}}\in\mathcal{C}$ with $\overline{\overline{58-c}}$. So, if we prefer to have all elements less than $\frac{p-1}{2}$, this is possible: $\left\{\overline{\overline{1}},\overline{\overline{9}},\overline{\overline{23}},\overline{\overline{25}},
\overline{\overline{7}},\overline{\overline{5}},\overline{\overline{13}}\right\}$ is a quasi-perfect code in $CAC^{e}(58,5)$.\\

$Example \ 4.$ Let $k=3$, $p=86413$. Choose $g=44659$. Then
$$\{ind_{g}(i) \; | \; i \ \in [1,12]\}=\{0, 81329, 63398, 76246, 76773, 58315, 72689, 71163, 40384,
71690, 73465, 53232\}$$
and so
$$\{ind_{g}(i) \ (\mbox{mod}\ 6) \; | \; i=1,3,\cdots,11\}=\{0, 2, 3, 5, 4, 1\}=[0,5],$$
$$\{ind_{g}(i) \ (\mbox{mod}\ 6) \; | \; i=2,4,\cdots,12\}=\{5, 4, 1, 3, 2, 0\}=[0,5].$$
Therefore, since $g^{6}\equiv 93989 \ (\mbox{mod}\ 172862),$ we obtain the following  quasi-perfect code in $CAC^{e}(172862,13)$:
$$\left\{\overline{\overline{93989^{j} \ (\mbox{mod}\ 172862)}}|0\leq j<7201\right\}.$$ \\



\begin{thebibliography}{1}
\vskip2mm
\bibitem{FLM} H. L. Fu, Y. H. Lin and M. Mishima, ``Optimal conflict-avoiding codes of even length and weight 3,'' \emph{IEEE Trans. Inf. Theory}, vol. 56, no. 11, pp. 5747--5756, Nov. 2010.
\bibitem{K} K. Momihara, ``Necessary and sufficient conditions for tight equi-difference conflict-avoiding codes of weight three,'' \emph{Des. Codes Cryptogr.}, vol. 45, no. 3, pp. 379--390, Dec. 2007.
\bibitem{MSJ} K. Momihara, J. Satoh and M. Jimbo, ``Constant weight conflict-avoiding codes,'' \emph{SIAM J. Discrete Math.}, vol. 21, no. 4, pp. 959--979, Jan. 2007.
\bibitem{KWC} K. W. Shum, W. S. Wong and C. S. Chen, ``A general upper bound on the size of constant-weight conflict-avoiding codes,''  \emph{IEEE Trans. Inf. Theory}, vol. 56, no. 7, pp. 3265--3276, Jul. 2010.
\bibitem{LI} L. Gy\"{o}rfi and I. Vajda,``Constructions of protocol sequences for multiple access collision channel without feedback,'' \emph{IEEE  Trans. Inf. Theory}, vol. 39 no. 5, pp. 1762--1765, Sep. 1993.
\bibitem{JMJ} M. Jimbo, M. Mishima, S. Janiszewski, A. Y. Teymorian and V. D. Tonchev, ``On conflict-avoiding codes of length $n=4m$ for three active users,'' \emph{IEEE Trans. Inf. Theory}, vol. 53, no. 8, pp. 2732--2742, Aug. 2007.
\bibitem{MFU} M. Mishima, H. L. Fu and S. Uruno, ``Optimal conflict-avoiding codes of length $n\equiv0 \ (\mbox{mod} \ 16)$ and weight 3,'' \emph{Des. Codes Cryptogr.}, vol. 52, no. 3, pp 275--291, Sep. 2009.
\bibitem{P} P. Mathys, ``A class of codes for a $T$ active users out of $N$ multiple-access communication system,'' \emph{IEEE Trans. Inf. Theory}, vol. 36, no. 6, pp. 1206--1219, Nov. 1990.
\bibitem{QLJ} Q. A. Nguyen, L. Gy\"{o}rfi and J. L. Massey, ``Constructions of binary constant-weight cyclic codes and cyclically permutable codes,'' \emph{IEEE Trans. Inf. Theory}, vol. 38, no. 3, pp. 940--949, May 1992.
\bibitem{WF} S. L. Wu and H. L. Fu, ``Optimal tight equi-difference conflict-avoiding codes of length $n=2k\pm1$ and weight 3,'' \emph{J.  Des. Comb.}, vol. 21, no. 6, pp. 223--231, Jun. 2013.
\bibitem{KLY} T. Kl{\o}ve, J. Luo and S. Yari, ``Codes correcting single errors of limited magnitude,'' \emph{IEEE Trans. Inf. Theory}, vol. 58, no. 4, pp. 2206--2219, Apr. 2012.
\bibitem{V} V. I. Levenshtein, ``Conflict-Avoiding Codes and Cyclic Triple Systems,'' \emph{Probl. Inf. Transm.}, vol. 43, no. 3, pp. 199--212, Sep. 2007.
\bibitem{LMJ} Y. Lin, M. Mishima and M. Jimbo, ``Optimal equi-difference conflict-avoiding codes of weight four,'' \emph{Des. Codes Cryptogr.}, vol. 78, no. 3, pp. 747--776, Mar. 2016.
\bibitem{WCD} W. Ma, C. E. Zhao and D. Shen, ``New optimal constructions of conflict-avoiding codes of odd length and weight 3,'' \emph{Des. Codes Cryptogr.}, vol. 73, no. 3, pp. 791--804, Dec. 2014.
\bibitem{LMS} Y. Lin, M. Mishima, J. Satoh and M. Jimbo, ``Optimal equi-difference conflict-avoiding codes of odd length and weight three,'' \emph{Finite Fields Their  Appl.}, vol. 72, no. 2, pp. 289--309, Aug. 2014.

\end{thebibliography}
\end{document}